\PassOptionsToPackage{table}{xcolor}
\documentclass[sigconf]{acmart}
\copyrightyear{2024}
\acmYear{2024}
\setcopyright{acmlicensed}\acmConference[ASE '24]{39th IEEE/ACM International Conference on Automated Software Engineering }{October 27-November 1, 2024}{Sacramento, CA, USA}
\acmBooktitle{39th IEEE/ACM International Conference on Automated Software Engineering (ASE '24), October 27-November 1, 2024, Sacramento, CA, USA}
\acmDOI{10.1145/3691620.3695529}
\acmISBN{979-8-4007-1248-7/24/10}

\AtBeginDocument{%
  }

\usepackage{pifont}
\usepackage{booktabs}
\usepackage{graphicx}
\usepackage{multirow}
\usepackage{multicol}
\usepackage{subcaption}
\usepackage{makecell}
\usepackage{adjustbox}
\usepackage{bm}
\usepackage{framed}
\usepackage{tcolorbox}
\usepackage{diagbox}
\usepackage{amsmath}
\usepackage{booktabs}
\usepackage{caption}
\usepackage{enumitem}
\usepackage[normalem]{ulem}
\usepackage{graphicx} 
\usepackage{xcolor}
\usepackage{color}
\useunder{\uline}{\ul}{}

\newcommand{\SlateBlue}[1]{\textcolor[RGB]{199,21,133}{#1}}
\newcommand{\yl}[1]{\SlateBlue{[Lin:#1]}}

\newcommand{\finding}[1]{ 
\begin{tcolorbox}[boxsep=2pt,left=3pt, right=3pt, top=2pt, bottom=2pt]
\textbf{Finding \refstepcounter{num}\thenum}: #1
\end{tcolorbox}
}

\newcommand{\ins}[1]{{\color{black}{#1}}}
\newcommand{\del}[1]{{\color{black}{}}}

\usepackage{listings}
\usepackage{color}

\definecolor{codegreen}{rgb}{0,0.6,0}
\definecolor{codegray}{rgb}{0.5,0.5,0.5}
\definecolor{codepurple}{rgb}{0.58,0,0.82}
\definecolor{backcolour}{rgb}{0.95,0.95,0.92}

\lstdefinestyle{mystyle}{
    backgroundcolor=\color{backcolour},   
    commentstyle=\color{codegreen},
    keywordstyle=\color{magenta},
    numberstyle=\tiny\color{codegray},
    stringstyle=\color{codepurple},
    basicstyle=\footnotesize,
    breakatwhitespace=false,         
    breaklines=true,                 
    captionpos=b,                    
    keepspaces=true,                 
    numbers=left,                    
    numbersep=5pt,                  
    showspaces=false,                
    showstringspaces=false,
    showtabs=false,                  
    tabsize=2
}

\lstdefinelanguage{Markdown}{
  basicstyle=\sffamily\footnotesize\color{black}, 
  moredelim=[s][\bfseries\itshape]{\{}{\}},
  commentstyle=\color{codegray},
  columns=fullflexible, 
  basewidth={0.6em,0.55em}, 
  morecomment=[l]{//}, 
  moredelim=**[is][\color{red}]{`}{`},
  moredelim=*[s][\color{red}]{```}{```}, 
  moredelim=**[is][\bfseries]{**}{**},
}

\begin{document}

\title{On the Evaluation of Large Language Models in Unit Test Generation}


\author{Lin Yang}
\authornote{Both authors contributed equally to this research.}
\authornote{This paper was completed during an internship at Huawei Cloud Computing Co. Ltd.}
\affiliation{
\institution{College of Intelligence and Computing, Tianjin University}
\country{China}
}
\email{linyang@tju.edu.cn}

\author{Chen Yang}
\authornotemark[1]
\affiliation{
\institution{College of Intelligence and Computing, Tianjin University}
\country{China}
}
\email{yangchenyc@tju.edu.cn}

\author{Shutao Gao}
\affiliation{
\institution{School of Future Technology, Tianjin University}
\country{China}
}
\email{gaoshutao@tju.edu.cn}

\author{Weijing Wang}
\affiliation{
\institution{College of Intelligence and Computing, Tianjin University}
\country{China}
}
\email{wangweijing@tju.edu.cn}

\author{Bo Wang}
\affiliation{
\institution{School of Computer and Information Technology, Beijing Jiaotong University}
\country{China}
}
\email{wangbo_cs@bjtu.edu.cn}

\author{Qihao Zhu}
\affiliation{
\institution{Key Laboratory of HCST, MoE DCST, Peking University}
\country{China}
}
\email{zhuqh@pku.edu.cn}

\author{Xiao Chu}
\affiliation{
\institution{Huawei Cloud Computing Co. Ltd.}
\country{China}
}
\email{chuxiao1@huawei.com}

\author{Jianyi Zhou}
\affiliation{
\institution{Huawei Cloud Computing Co. Ltd.}
\country{China}
}
\email{zhoujianyi2@huawei.com}

\author{Guangtai Liang}
\affiliation{
\institution{Huawei Cloud Computing \\Co. Ltd.}
\country{China}
}
\email{liangguangtai@huawei.com}

\author{Qianxiang Wang}
\affiliation{
\institution{Huawei Cloud Computing \\Co. Ltd.}
\country{China}
}
\email{wangqianxiang@huawei.com}

\author{Junjie Chen}
\authornote{Corresponding Author}
\affiliation{
\institution{College of Intelligence and Computing, Tianjin University}
\country{China}
}
\email{junjiechen@tju.edu.cn}

\renewcommand{\shortauthors}{Yang et al.}

\begin{abstract}

Unit testing is an essential activity in software development for verifying the correctness of software components. However, manually writing unit tests is challenging and time-consuming. The emergence of Large Language Models (LLMs) offers a new direction for automating unit test generation. Existing research primarily focuses on closed-source LLMs (e.g., ChatGPT and CodeX) with fixed prompting strategies, leaving the capabilities of advanced open-source LLMs with various prompting settings unexplored. Particularly, open-source LLMs offer advantages in data privacy protection and have demonstrated superior performance in some tasks. Moreover, effective prompting is crucial for maximizing LLMs' capabilities.
In this paper, we conduct the first empirical study to fill this gap, based on 17 Java projects, five widely-used open-source LLMs with different structures and parameter sizes, and comprehensive evaluation metrics. Our findings highlight the significant influence of various prompt factors, show the performance of open-source LLMs compared to the commercial GPT-4 and the traditional Evosuite, and identify limitations in LLM-based unit test generation. We then derive a series of implications from our study to guide future research and practical use of LLM-based unit test generation.

\end{abstract}



\maketitle

\section{Introduction}
Unit testing focuses on verifying the functionality of each individual program component (such as each method) in order to ensure it works as intended.
Hence, writing high-quality unit tests is crucial, which facilitates developers to catch defects early and diagnose them efficiently.
However, manually writing these tests is time-consuming and tedious, which could significantly escalate the cost of software development~\cite{kumar2016impacts}.
To reduce this effort, several approaches have been proposed to automatically generate unit tests. 
Traditional approaches adopted symbolic execution~\cite{DBLP:conf/issta/PasareanuMBGLPP08,DBLP:conf/tacas/XieMSN05}, evolutionary algorithms~\cite{evosuite}, or model checking~\cite{DBLP:journals/sttt/EnoiuCOWSP16,DBLP:conf/esec/GargantiniH99} to automatically generate unit tests.
While they can generate tests with promising coverage, they still fall short of the high utility achieved by human-written tests.
For example, the tests can be difficult to maintain and understand due to the code smell issue, such as lack of meaningful identifiers~\cite{palomba2016diffusion}. 
To alleviate this problem, several deep-learning-based (DL-based) approaches have been proposed \cite{a3test, athenatest}. 
They collected a large corpus of unit tests with their corresponding focal method to build DL models, thereby enhancing their test generation effectiveness.
However, due to limitations in the scale of models and training data, they often struggle to fully understand the code's intent and generate high-quality tests.

Although existing approaches have made significant progress, automatically generating unit tests still faces the following challenges:
(1) \textit{Generating syntactically correct tests.}
Unit test generation can be treated as a sub-problem of code generation, and thus also suffers from the general problem of ensuring that the generated code conforms to language's grammar rules.
(2) \textit{Generating effective tests.}
High-quality unit tests should thoroughly explore the behaviour of the target program component, which requires the approach to understand code intent and structures and thus generate tests with high test coverage and defect detection ability.
(3) \textit{Generating maintainable tests.}
In software evolution, we prefer unit tests that are easy to read and change, which requires the approach to generate tests following the code practice and smell well.

Recently, Large Language Models (LLMs), with supreme abilities in code understanding and natural language processing, have shown potential in code generation tasks.
Researchers also proposed to leverage the most popular LLMs, i.e., GPT-3.5 or GPT-4, to automatically generate unit tests~\cite{chatunitest,chattester,exploreingllminutg}.
However, these LLMs are commercial and closed-source, which presents practical challenges due to concerns over data security and the costs associated with API usage.
Additionally, these existing studies predominantly relied on fixed prompting strategies based on prior experience, neglecting the potential influence of various prompting factors, such as prompt design choices and in-context learning (ICL) methods. 
As shown in existing studies, prompting is crucial in maximizing the capabilities of LLMs~\cite{autoCoT,CoT_scaling,structuredCoT,logicot}, and thus this lack of in-depth analysis on these factors could significantly limit the development of actionable guidelines for optimizing unit test generation with LLMs~\cite{siddiq2024quality,ICLsurvey}.
Moreover, with the rapid emergence of open-source code LLMs and various prompting methods, their effectiveness for unit test generation remains largely unexplored.
Therefore, it is urgent to investigate the effectiveness of advanced open-source LLMs in unit test generation with various prompting methods.

In this work, we conducted the first extensive study to achieve the above-mentioned goal based on 17 Java projects from the Defects4J 2.0 benchmark~\cite{defects4j}.
Particularly, we evaluated five open-source code LLMs, which are built on top of the widely-used CodeLlama~\cite{codellama} and DeepSeek-Coder~\cite{deepseek-coder} structures with diverse model scales ranging from 7B to 34B.
Based on the extensive experiments under the guidance of our elaborately-designed four research questions (RQs), we obtain a series findings about LLM-based unit test generation and deliver the corresponding actionable guidelines for future research and practice use in this field, which are mainly summarized as follows:
\begin{itemize}[leftmargin=0.7cm]
    \item[(1)] The prompt design (the description style and selected code features) is crucial to the effectiveness of LLMs in unit test generation. It is recommended to align the description style with the training data and choose code features considering the LLMs' code comprehension ability and the space left for generating unit tests. Notably, including other methods defined in the target class (except the focal method) negatively impacts the overall effectiveness of LLM-based unit test generation due to their extensive length.
    
    \item[(2)]  The conclusions drawn from open-source LLMs in other tasks do not necessarily generalize to unit test generation, including dominance relationships among studied LLMs. However, all studied LLMs, including the state-of-the-art GPT-4, underperform traditional Evosuite in terms of test coverage. This is primarily due to the large percentage of syntactically invalid unit tests generated by LLMs, a result of LLMs' hallucination. Therefore, effective solutions are needed, e.g., designing post-processing rules to fix common syntactic issues.

    \item [(3)] Despite their effectiveness in other tasks, directly adapting the Chain-of-Thoughts (CoT) and Retrieval Augmented Generation (RAG) methods for unit test generation does not improve effectiveness and may even reduce it in some cases. CoT is primarily limited by the LLMs' code comprehension ability, while RAG is constrained by the significant gap between the retrieved unit tests and those that LLMs excel at generating. Special design for the use of ICL methods in unit test generation is required.
    \item[(4)] The defect detection ability of LLM-generated unit tests is limited, primarily due to their low validity. Although valid unit tests are generated by LLMs for many defects, a significant number of defects remain undetected, mainly because the tests fail to produce the specific inputs necessary to trigger these defects. Therefore, designing effective mutation strategies for the inputs within generated unit tests could further improve defect detection effectiveness.
\end{itemize}

In summary, this paper makes the following contributions:

\begin{itemize}[leftmargin=0.7cm]
\item We performed the first empirical study to evaluate open-source LLMs in unit test generation. We studied five powerful code LLMs, ranging from 7B to 34B parameters, on 17 Java projects from the well-known Defects4J benchmark. Our experiments required around 3,000 NVIDIA A100 GPU-hours, underscoring the scale and intensity of our study in advancing LLM-based unit test generation.

\item We evaluated LLM-based unit test generation from four aspects defined by our research questions, including the influence of prompt design, the influence of ICL methods, the comparison among open-source LLMs, the commercial GPT-4, and traditional Evosuite, and the evaluation in terms of diverse metrics (including syntactic validity, test coverage, and defect detection).


\item We summarized nine major findings from our extensive study and delivers a series of implications for guiding future research and practice use of LLM-based unit test generation.

\end{itemize}

\section{Study Design}

In general, when using LLMs to address some code-related task, it involves three main aspects: an LLM, a prompt as the input of the LLM, and an in-context (ICL) method employed to improve LLMs in addressing the task, and thus it is important to investigate these aspects in LLM-based unit test generation. 
Furthermore, as the goal of unit testing is to detect defects at an early stage, the defect detection ability of the generated unit tests by LLMs is also crucial.
In this work, to conduct a comprehensive study, the workflow of our study consists of four parts that target each aspect of LLM-based unit test generation as discussed above. 

\subsection{Research Questions}

We investigated the unit test generation effectiveness of LLMs by answering the following research questions (RQs):

\begin{itemize}[leftmargin=0.4cm]
    \item \textbf{RQ1}: How does prompt design affect the effectiveness of LLMs in unit test generation?

    \item \textbf{RQ2}: How do open-source LLMs perform in unit test generation compared to GPT-4 and Evosuite?

    \item \textbf{RQ3}: How do in-context learning methods affect the effectiveness of LLM-based unit test generation?

    \item \textbf{RQ4}: How effective are the unit tests generated by LLMs in terms of defect detection?
\end{itemize}

In RQ1, we investigated the effectiveness of each studied open-source LLM in unit test generation by designing different prompts.
This helps investigate the influence of different prompt designs on this task and also helps determine how to design a prompt to make LLMs achieve the best effectiveness.
In RQ2, we compared the studied open-source LLMs in unit test generation based on the best prompt design we explored in RQ1, and employed the state-of-the-art commercial LLM (i.e., GPT-4) and the widely-studied traditional approach (i.e., Evosuite) as reference.
This helps understand the current situation of these open-source LLMs in unit test generation.
In RQ3, we incorporated two widely-used ICL methods to improve the usage of these studied LLMs respectively, and then revisited the effectiveness of these LLMs in unit test generation.
This is helpful to explore whether the state-of-the-art ICL methods in general tasks can help improve the effectiveness of LLMs in the specific task of unit test generation.

The former three RQs correspond to the above-mentioned three aspects affecting the effectiveness of LLMs in addressing some tasks, which adopted the widely-used evaluation metrics in existing studies on LLM-based unit test generation (i.e., syntactic validity and test coverage).
RQ4 further complements them by considering another key ability (i.e., the defect detection ability) of unit tests.
That is, in RQ4, we investigated the effectiveness of these studied LLMs in terms of fault detection with the corresponding automatically generated unit tests.

\subsection{Studied LLMs}
\label{sec:LLMs}

First, we selected a set of representative code LLMs for investigation in our study.
Our selection process started with the leaderboard of code LLMs hosted on Hugging Face~\cite{bigcode-leaderboard}. 
We first sorted code LLMs based on their scores and found that the most powerful LLMs shared two structures: CodeLlama and DeepSeekCoder. 
Then, we determined a set of code LLMs with different scales for each structure.
They are CodeLlama-7B-Instruct, CodeLlama-13B-Instruct, Phind-CodeLlama-34B-v2, DeepSeekCoder-6.7B-Instruct, and DeepSeekCoder-33B-Instruct.
Besides, we used the state-of-the-art commercial LLM (i.e., GPT-4~\cite{gpt4}), as a reference due to its well-known effectiveness. 



\textbf{CodeLlama} refers to a series of code-specific LLMs evolved from Llama 2~\cite{llama}, available in three sizes (i.e., 7B, 13B, and 34B) and three variants (i.e., base model, Python fine-tuned and instruction-tuned models). 
\textbf{Phind} models are the fine-tuned versions of CodeLlama-34B, leveraging an internal Phind corpus and instruction-tuning, and they have outperformed other models in CodeLlama series. 
In our study, we selected three instruction-tuned LLMs: CodeLlama-7B-Instruct, CodeLlama-13B-Instruct, and Phind-CodeLlama-34B-v2, as instruction-tuned models tend to exhibit superior performance as shown in the existing work~\cite{yuan2023evaluating}. 
For ease of presentation, we call them \textbf{CL-7B}, \textbf{CL-13B}, and \textbf{PD-34B}, respectively.

\textbf{DeepSeekCoder} comprises a series of code LLMs trained on a repository-level code corpus, including both pre-trained and instruction-tuned models. These models come in various sizes, including 1.3B, 5.7B, 6.7B, and 33B. 
To ensure fair comparisons with our selected CodeLlama models, we selected DeepSeekCoder-6.7B-Instruct and DeepSeekCoder-33B-Instruct since they have comparable parameter sizes with CL-7B and PD-34B.
We call them \textbf{DC-7B} and \textbf{DC-33B}, respectively.

\textbf{ChatGPT} includes influential commercial closed-source LLMs developed by OpenAI, such as GPT-3.5 and GPT-4. These models are pre-trained on trillions of tokens and further enhanced through instruction-tuning and reinforcement learning with human feedback~\cite{RLHF}. 
They have shown exceptional performance in various tasks~\cite{chattester,chatunitest}. 
Among the ChatGPT series, GPT-4 is the most advanced and outperforms its predecessors. Therefore, we selected \textbf{GPT-4} as the representative closed-source LLM for reference.
\subsection{Prompt Design}
\label{sec:prompt_design}


Prompts are crucial for an LLM as they contain the context, instruct the task, and largely influence the model's performance~\cite{prompt-tuning,DBLP:journals/corr/abs-2403-03028,DBLP:journals/corr/abs-2401-14423,ICLsurvey}.
In general, when handling code-related tasks, the prompt should include proper code features, and the task and these code features should be described in a proper style, which can facilitate the LLM understanding and thus address this task.
That is, when designing a prompt, we should take two factors into consideration:
(1) which style is used to describe the task and these code features, and (2) which code features should be included.

\textbf{Description Style.}
There are two widely-used description styles in addressing code-related tasks with LLMs in the literature~\cite{chattester,chatunitest,TestPilot,mutap}: \textit{natural language description}~\cite{chatunitest,chattester} and \textit{code language description}~\cite{TestPilot,exploreingllminutg,mutap}.
The former describes the task and code features in natural language, while the latter makes the prompt a code fragment by introducing the task and code features in corresponding comments.
In our study, we investigated the influence of the two styles on the effectiveness of each studied open-source LLM in unit test generation.

\textbf{Code Features.}
To adequately investigate the influence of different code features, we surveyed the existing studies on LLM-based unit test generation~\cite{chattester,a3test,chatunitest,mutap,exploreingllminutg}, and collected all their used code features.
In total, we collected six code features and categorized them into three types:
    (1) \textit{Focal Method} is the target method to be tested. The code features extracted from it include the method body ($\textit{FM}_b$) and its parameters ($\textit{FM}_{p}$).
    (2) \textit{Focal Class} is the class containing the focal method. 
    The code features extracted from it include its constructor ($\textit{FC}_{c}$), the fields defined within the class ($\textit{FC}_{f}$), and other methods defined within the class ($\textit{FC}_{m}$).
    (3) \textit{Related Classes} are the classes including the constructors of the parameters in the focal method (except the focal class).
    The code feature extracted from it is these constructors ($\textit{RC}_{c}$).
Among them, $\textit{FM}_b$ is the basic content in the prompt for LLM-based unit test generation.
Thus we performed an ablation experiment to investigate the contribution of each of the remaining five code features.



\subsection{In-Context Learning Method}
\label{sec:icl}
\textit{Chain-of-Thoughts} (CoT)~\cite{chattester} and \textit{Retrieval Augmented Generation} (RAG)~\cite{rag} are two widely-used ICL methods in various code-related tasks.
In this study, we adapted them to unit test generation task, and investigated the impact of these methods on the effectiveness of unit test generation. 

\textbf{CoT} was originally proposed to improve the reasoning ability of LLMs in addressing mathematical and common-sense reasoning tasks~\cite{CoT_scaling}.
Subsequently, it has been applied to some code-related tasks and the effectiveness has been also demonstrated through extensive experiments~\cite{chattester,structuredCoT}.
Particularly, the existing work has demonstrated the effectiveness of applying CoT to ChatGPT for unit test generation~\cite{chattester}, which instructs ChatGPT to (1) understand the focal method and then (2) generate unit tests according to its understanding.
However, it is unclear whether CoT is still effective in improving open-source LLMs in unit test generation.
To fill the gap, this work performs the first study that extensively investigates the effectiveness of COT on various widely used open-source LLMs.
\textbf{RAG} enhances LLMs by incorporating relevant information retrieved from external databases to improve prompt contents, which has been widely used in code generation~\cite{REDCODER,rag2,rag3}.
In general, RAG first retrieves similar inputs, and then treats these inputs along with their corresponding ground-truth outputs as examples to enhance LLMs.
In our study, we investigated its effectiveness in improving LLM-based unit test generation through adaptation.
Specifically, it first retrieves the most similar method to the focal method within the project.
Following the existing work~\cite{rag3, DPR}, we used CodeBERT to encode the semantics of methods as vectors and then measured the cosine similarity between vectors.
Then, the retrieved method, along with its corresponding unit tests, was integrated into the prompt to instruct LLMs to generate unit tests.

\subsection{Benchmark}
\label{sec:data_processing}
We used the Defects4J 2.0 benchmark~\cite{defects4j} to evaluate the effectiveness of LLMs in unit test generation.
It consists of 835 real-world defects collected from 17 open-source Java projects and has been widely used in software testing~\cite{kang2023large}, fault localization~\cite{jiang2023variable,jiang2019combining} and program repair~\cite{10.1145/3213846.3213871, wen2018context,liu2019tbar, d4jtestset}.
In this benchmark, each defect corresponds to a buggy version and a fixed version.
The fixed version indicates that the patches have been applied to certain methods and then the defect can be fixed.
To balance the evaluation costs and defect detection capability investigation, we regarded each \textit{patched} method as a focal method to be tested, instead of all methods in this benchmark.
Particularly, following the testing practice and the existing work~\cite{a3test,athenatest}, we just considered \textit{public} methods as focal methods.
For a bug involving multiple buggy methods, we used all the \textit{public} methods as focal methods.
In total, we collected 778 focal methods, which involve 413 defects from all 17 projects in the benchmark.
Due to the space limit, we put the numbers of focal methods and involved defects for each project at our homepage~\cite{homepage}.

\subsection{Implementation}
\label{sec:post_processing}
Due to the variable inconsistent output format of LLMs, we first used the widely-used AST parser (i.e., tree-sitter\cite{tree-sitter}) to extract the generated unit tests from the outputs, rather than relying on simple text matching.
Then, we integrated all test methods into a test class for subsequent evaluation.
Finally, we imported the classes within the project and some common dependencies required by JDK (e.g., the {\tt java.utils} package) and JUnit (e.g., the {\tt org.junit.Assert} package) into the test class, in order to avoid simple compilation issues caused by missing dependencies. 

We developed our experimental scripts based on PyTorch 2.0.0~\cite{torch} and transformers 4.34.1~\cite{transformers}, and used VLLM~\cite{vllm} libraries to speed up our experiments. 
All the experiments were conducted on four servers with the same configuration, i.e., Ubuntu 18.04 LTS, Intel Xeon Gold 6240C CPU, 512GB RAM, and eight NVIDIA A100 GPUs. 
All of our code and data are available at our project homepage~\cite{homepage}.

\subsection{Metrics}
Following the existing studies~\cite{chattester, chatunitest}, we leveraged the three widely-used metrics to evaluate the generated unit tests by each technique in RQ1, RQ2, and RQ3: 
\textbf{Compilation Success Rate} (\textbf{CSR}), \textbf{Line Coverage} (\textbf{$Cov_L$}), and \textbf{Branch Coverage} (\textbf{$Cov_B$}).
As introduced in Section~\ref{sec:post_processing}, for each focal method, we integrated all the generated unit tests into a test class, and then compiled the test class.
Therefore, we calculated CSR by the ratio of the number of test classes compiled successfully over the total number of test classes corresponding to all focal methods.
Then, we executed each test class that was successfully compiled, and collected its achieved line coverage and branch coverage via the code coverage collection tool (i.e., Jacoco~\cite{jacoco}).
For each test class with compilation errors, we recursively remove test methods occurred in compilation error messages from the test class and re-execute the test class until it can be compiled successfully or there is no test method left in the test class.
In this way, we try our best to collect as much coverage achieved as possible.
Higher CSR, $Cov_L$, and $Cov_B$ indicate better effectiveness in unit test generation.

Furthermore, we incorporated an additional metric in RQ4, i.e., \textbf{the Number of Detected Defects} (\textbf{NDD}), to evaluate the defect detection ability of the generated unit tests.
Specifically, if a test class passes on a fixed version but fails on the corresponding buggy version, it indicates that the test class detected a defect. 

\section{Results}
\newcounter{num}

\subsection{Influence of Prompt Design}
\label{sec:rq1}
\begin{table*}[t]
\centering
\caption{Effectiveness of Open-source LLMs Using Different Description Styles in Prompt.}
\label{tab:rq1_prompt_strategy}
\adjustbox{}{%
\begin{tabular}{@{}c|ccc|ccc|ccc|ccc|ccc@{}}
\toprule
\multirow{2}{*}{\textbf{Style}} &
  \multicolumn{3}{c|}{\textbf{CL-7B}} &
  \multicolumn{3}{c|}{\textbf{CL-13B}} &
  \multicolumn{3}{c|}{\textbf{PD-34B}} &
  \multicolumn{3}{c|}{\textbf{DC-7B}} &
  \multicolumn{3}{c}{\textbf{DC-33B}} \\  
 &
  \textbf{$\bm{CSR}$} &
  \textbf{$\bm{Cov_{L}}$} &
  \textbf{$\bm{Cov_{B}}$} &
  \textbf{$\bm{CSR}$} &
  \textbf{$\bm{Cov_{L}}$} &
  \textbf{$\bm{Cov_{B}}$} &
  \textbf{$\bm{CSR}$} &
  \textbf{$\bm{Cov_{L}}$} &
  \textbf{$\bm{Cov_{B}}$} &
  \textbf{$\bm{CSR}$} &
  \textbf{$\bm{Cov_{L}}$} &
  \textbf{$\bm{Cov_{B}}$} &
  \textbf{$\bm{CSR}$} &
  \textbf{$\bm{Cov_{L}}$} &
  \textbf{$\bm{Cov_{B}}$} \\ \midrule
NL &
  \textbf{40.31} &
  \textbf{18.76} &
  \textbf{16.67} &
  \textbf{53.86} &
  \textbf{18.55} &
  \textbf{17.11} &
  51.93 &
  28.59 &
  23.51 &
  47.36 &
  23.79 &
  18.66 &
  49.07 &
  26.98 &
  23.91 \\
CL &
  34.37 &
  16.21 &
  14.34 &
  38.95 &
  14.16 &
  12.18 &
  \textbf{56.06} &
  30.31 &
  24.14 &
  50.27 &
  23.94 &
  17.20 &
  50.54 &
  \textbf{32.36} &
  26.76  \\ 
  
  \bottomrule
\end{tabular}%
}
\end{table*}
\textbf{Influence of Description Style.}
We first investigated the influence of description styles, i.e., natural language description (NL)  and code language description (CL).
Specifically, we investigated the effectiveness of each studied open-source LLM by incorporating all studied code features into the prompt in each studied description style, respectively.
Table~\ref{tab:rq1_prompt_strategy} shows the results in terms of CSR, $Cov_{L}$, and $Cov_{B}$.
To evaluate the difference between NL and CL in statistics, we performed the Wilcoxon rank sum test~\cite{wilcoxon_rank_sum_test} with a significance level of 0.05 between them and calculated the Rank-biserial correlation~\cite{pallant2020spss} scores to show the effect size. 
If a \textit{p-value} is smaller than 0.05, it indicates that there is a statistically significant difference between NL and CL for the used LLM in terms of the metric. 
An \textit{effect-size value} above 0.3 generally indicates a meaningful difference between compared groups~\cite{cohen2013statistical}. 
In this table, we used bold to highlight the better one with statistical and meaningful significance according to the p-value (<0.05) and the effect size (>0.3) between NL and CL.


From Table~\ref{tab:rq1_prompt_strategy}, CL-7B and CL-13B perform better when using the NL style than using the CL style in terms of all three used metrics.
In particular, the superiority of NL over CL is statistically significant in all six cases for CL-7B and CL-13B.
For example, the unit tests generated by CL-7B in the NL style achieve 2.55\% and 2.33\% higher line coverage and branch coverage than those in the CL style.
For DC-7B, DC-33B, and PD-34B, there is no statistically significant difference between the two styles in most (seven out of nine) cases, even though the CL style performs slightly better than the NL style in general.
The results indicate that DC-7B, DC-33B, and PD-34B are more robust with respect to the description style used in prompt than CL-7B and CL-13B.



We then investigated the reason behind the above phenomenon and found that the root cause lies in which description styles are employed in the training data for LLMs.
Specifically, CL-7B and CL-13B were build based on Llama2 by using a code corpus with 500B tokens. Although they have learned how to understand code language, the base Llama2 model still have dominant impact on their ability of understanding prompts.
In contrast, although PD-34B shares the Llama2 architecture, it conducted three rounds of additional training with CL data (i.e., from Llama2 to CodeLlama-34B, then Phind-v1 and v2), and thus it performs stably regardless of the used styles.
Similarly, DC-7B and DC-33B were initially trained on a hybrid corpus comprising 80\% data in the CL style and 20\% data in the NL style, allowing them to work well with both styles.

Overall, when utilizing an LLM to address the task of unit test generation, it is preferable to design the prompt in the description style that the LLM was trained on, which is usually overlooked in practice.
For sufficient comparisons, we adopted the style making the LLM perform better for each LLM in the following experiments.
Specifically, we designed the prompt in the NL style for CL-7B and CL-13B and in the CL style for DC-7B, DC-33B, and PD-34B.
\finding{CL-7B and CL-13 perform significantly better when using the style of NL to instruct them, while PD-34B, DC-7B, and DC-33B perform stably regardless of the used styles. The underlying reason lies in whether the description styles used for instructing LLMs are well aligned with those used in the training data for LLMs.}

\label{sec:ablation_study_features}
\begin{table}[t]
\centering
\caption{Effectiveness of Unit Tests Generated By Open-Source LLMs with Different Prompt Variants in Terms of Compilation Success Rate (CSR).}
\label{tab:rq2_compilation_success_rate}
\resizebox{\columnwidth}{!}{
\begin{tabular}{l|rrrrr}
\toprule
\multicolumn{1}{c|}{\textbf{Variants}} & \textbf{CL-7B} & \textbf{CL-13B} & \textbf{PD-34B}        & \textbf{DC-7B} & \textbf{DC-33B}    

\\ \midrule
\multicolumn{1}{c|}{\textbf{All}}                                                    & 40.31 & 53.86  & 55.49  & 49.22  & 45.65 
\\ 
\midrule








$\bm{w/o}$ $\bm{FC_{c}}$ & {+0.99} & {-1.87} & {-0.42} & {-1.00} & {-2.43} 

\\
$\bm{w/o}$ $\bm{FC_{f}}$             & {+0.28} & {-3.43}  & {+0.29}  & {-0.86}  & {-0.29}

\\
$\bm{w/o}$ $\bm{FC_{m}}$           & {-2.14} & {\textbf{-14.92}} & {\textbf{-6.42}}  & {\textbf{-11.42}} & {\textbf{-4.28}}

\\ 

$\bm{w/o}$ $\bm{FM_{p}}$  & {+3.99} & {-1.01} & {+1.57} & {+1.84} & {-0.71} 

\\

 $\bm{w/o}$ $\bm{RC_{c}}$     & {+3.99} & {-1.01}  & {+1.57}  & {+1.84}  & {-0.71} 

\\ \bottomrule
\end{tabular}
}
\end{table}
\begin{table*}[t]

\caption{Effectiveness of Unit Tests Generated By Different Prompt Variants in Terms of Test Coverage.}
\label{tab:rq2_coverage_scores}
\adjustbox{max width=\textwidth}{%
\begin{tabular}{c|cc|cc|cc|cc|cc}
\toprule
\multicolumn{1}{c|}{\multirow{2}{*}{\textbf{Variants}}} &
  \multicolumn{2}{c|}{\textbf{CL-7B}} &
  \multicolumn{2}{c|}{\textbf{CL-13B}} &
  \multicolumn{2}{c|}{\textbf{PD-34B}} &
  \multicolumn{2}{c|}{\textbf{DC-7B}} &
  \multicolumn{2}{c}{\textbf{DC-33B}} 
  \\
\multicolumn{1}{c|}{} &
  $\bm{Cov_{L}}$ &
  $\bm{Cov_{B}}$ &
  $\bm{Cov_{L}}$ &
  $\bm{Cov_{B}}$ &
  $\bm{Cov_{L}}$ &
  $\bm{Cov_{B}}$ &
  $\bm{Cov_{L}}$ &
  $\bm{Cov_{B}}$ &
  $\bm{Cov_{L}}$ &
  $\bm{Cov_{B}}$ 
  \\ \midrule
\multicolumn{1}{c|}{\textbf{All}} &
  18.76 &
  16.67 &
  18.55 &
  17.11 &
  33.41 &
  27.12 &
  29.02 &
  23.77 &
  32.72 &
  29.26 
  \\ \midrule
  $\bm{w/o}$ $\bm{FC_{c}}$ &
  {\textbf{+8.09}} &
  {\textbf{+7.66}} &
  {-2.44} &
  {-0.12} &
  {+0.62} &
  {+1.62} &
  {-2.72} &
  {-2.01} &
  {-3.16} &
  {-2.31} 
  \\
  $\bm{w/o}$ $\bm{FC_{f}}$  &
  {\textbf{+7.46}} &
  {\textbf{+6.52}} &
  {+1.83} &
  {+0.92} &
  {\textbf{+5.17}} &
  {\textbf{+4.61}} &
  {-0.37} &
  {-0.75} &
  {+1.11} &
  {-0.70} 
  \\
  $\bm{w/o}$ $\bm{FC_{m}}$ &
  {\textbf{+9.26}} &
  {\textbf{+8.67}} &
  {\textbf{+3.07}} &
  {\textbf{+3.62}} &
  {\textbf{+4.04}} &
  {\textbf{+5.23}} &
  {-0.65} &
  {+2.27} &
  {+1.79} &
  {+0.05} 
  \\ 
  $\bm{w/o}$ $\bm{FM_{p}}$ &
  {\textbf{+8.93}} &
  {\textbf{+8.09}} &
  {-0.32} &
  {+0.37} &
  {+2.46} &
  {+2.57} &
  {-1.31} &
  {-1.18} &
  {-1.63} &
  {-1.91} 

  \\
  $\bm{w/o}$ $\bm{RC_{c}}$ &
  {\textbf{+8.96}} &
  {\textbf{+8.11}} &
  {-0.32} &
  {+0.37} &
  {+2.45} &
  {+2.56} &
  {-1.31} &
  {-1.18} &
  {\textbf{-1.90}} &
  {\textbf{-2.10}} 
  \\ \bottomrule
\end{tabular}
}
\end{table*}

\noindent\textbf{Influence of Code Features.}
We performed an ablation experiment to investigate the influence of code features (i.e., $FM_p$, $FC_c$, $FC_f$, $FC_m$, and $RC_c$ presented in Section~\ref{sec:prompt_design}).
Specifically, we created five variants for prompt design, each of which removes a code feature from the prompt incorporating all studied code features, respectively.
Then, we investigated the effectiveness of each studied open-source LLM with each of the prompt variants in its description style determined before.
Table~\ref{tab:rq2_compilation_success_rate} shows the results in terms of CSR and Table~\ref{tab:rq2_coverage_scores} shows the results in terms of $Cov_{L}$ and $Cov_{B}$.
In both tables, Rows ``All'' present the results of each studied LLM with the prompt incorporating all studied code features, and the remaining rows present the increased/decreased effectiveness over the corresponding effectiveness shown in Rows ``All'' for each studied LLM.
We also used bold to highlight the increment/decrement with statistical significance (\textit{p-values} smaller than 0.05 and \textit{effect size} greater than 0.3).

From Table~\ref{tab:rq2_compilation_success_rate}, all the cases with statistical significance are at Row ``w/o $FC_m$'' and all of them are decrement (ranging from 2.14\% to 14.92\%) in terms of CSR.
The results indicate the significant contribution of $FC_m$ (other methods defined within the focal class) to generate syntactic valid unit tests with LLMs.
The reason may be that building a unit test scenario often requires several steps and some necessary information exists in $FC_m$.
Hence, removing $FC_m$ may result in significant hallucination issues for LLMs, harming the validity of generated unit tests.
The remaining code features have no statistically significant influence on LLM-based unit test generation in terms of CSR, which is also as expected.
Specifically, (1) regarding $FC_f$, fields are seldom directly accessed from unit tests; (2) regarding $FC_c$ and $RC_c$, LLMs typically use no-parameter constructors for instantiation or employ mocked instances instead (e.g., about 58.49\% of generated test classes involve such instantiation methods on average); (3) the parameters of the focal method also occur in the method body (i.e., the basic content $FM_b$ that is always retained in the prompt).

From Table~\ref{tab:rq2_coverage_scores}, we found a surprising phenomenon, i.e., almost all cases at Row ``w/o $FC_m$'' are increment, even with statistical significance.
This indicates that removing $FC_m$ in prompt design is helpful to improve code coverage even though it is harmful to the syntactic validity of generated unit tests.
We carefully analyzed the underlying reason and found that the number of generated unit tests may be the key factor.
Specifically, these open-source LLMs have fixed-size space to include prompt contents and LLMs' response.
If the prompt is longer, it indicates that the contents of the response from the LLM are fewer, potentially leading to a smaller number of generated unit tests.
After removing $FC_m$ in prompt design, the window space left for the LLM's response becomes larger, and indeed, the average number of generated unit tests across all the studied open-source LLMs are significantly increased from 3,654 to 5,434, which is indeed helpful to improve code coverage.
Overall, this benefit even outperforms the negative influence on the syntactic validity of generated unit tests.

\finding{$FC_m$ contributes the most to the syntactic validity of generated unit tests among all studied code features, but its extensive length limits the number of generated unit tests with LLMs. The outperforming influence of the latter over the former finally makes $FC_m$ limit test coverage.}

Except $FC_m$, removing each of the remaining four code features is significantly helpful to improve code coverage for the three LLMs based on CodeLlama in general, but has no statistical significant influence on the two LLMs based on DeepSeek-Coder in terms of code coverage.
The underlying reason may lie in the ability of the foundation models in comprehending code features.
Specifically, the CodeLlama models are trained from Llama2, which are pre-trained on natural language data, while the DeepSeek-Coder models are built from scratch based on a corpus primarily composed of code language.
Therefore, DC-7B and DC-33B are able to comprehend provided code features more sufficiently, facilitating generating effective unit tests.
Regarding CL-7B, CL-13B, and PD-34B, because of the relatively weak ability in comprehending code features, the negative influence of missing some information due to removing code features is smaller than the brought benefit (i.e., leaving larger space for LLMs' response), eventually bringing the improvement of test coverage.
Additionally, the possible reason why removing $FC_m$ has positive influence on both CodeLlama and DeepSeek-Coder models is that $FC_m$ is significantly longer than the other code features, and thus the benefit in leaving space for LLMs' response due to its removal is more significant to test coverage improvement by generating more unit tests.
Overall, there seems to be a trade-off between including more code features and leaving space for generating more unit tests in prompt design.

Particularly, when incorporating all studied code features, the relatively small CodeLlama models (i.e., CL-7B and CL-13B) perform worse than the relatively small DeepSeek-Coder model (i.e., DC-7B) in terms of code coverage, and the relatively large CodeLlama model (i.e., PD-34B) performs closely to the relatively large DeepSeek-Coder model (i.e., DC-33B).
However, in our ablation experiment, the best effectiveness of CL-7B and CL-13B in terms of code coverage (i.e., that achieved by the prompt variant removing $FC_m$) is even better than that of DC-7B.
Similarly, the achieved code coverage with PD-34B using the prompt variant removing $FC_m$ outperforms the best effectiveness of DC-33B.
This indicates the importance of selecting code features for a given LLM due to the diverse characteristics of LLMs.
For sufficient comparisons, we adopted the prompt variant making the LLM perform the best for each LLM in the following experiments.
Specifically, we incorporated all studied code features in the prompt for DC-7B and DC-33B but removed $FC_m$ for the remaining three LLMs.
Note that the prompt variant performing the best in our ablation experiment may be not the global optimum, since we did not evaluate all combinations of these code features due to significant costs.
We will discuss this threat in detail in Section~\ref{sec:threats}.

\finding{Code features (except $FC_m$) have different influence on LLMs due to different abilities of their foundation models in comprehending code features. 
Using proper code features may largely enhance unit test generation effectiveness of LLMs (especially CodeLlama models).}

\subsection{Effectiveness Comparison} 
\label{sec:rq2}
\begin{table}[t]
\centering
\caption{Effectiveness Comparison of Studied LLMs and Evosuite in Unit Test Generation.}
\label{tab:rq1}
\adjustbox{max width=\columnwidth}{
\begin{tabular}{@{}cccccc@{}}
\toprule
\textbf{Model} & $\bm{CSR}(\%)$ & $\bm{Cov}_{L}(\%)$ & $\bm{Cov}_{B}(\%)$ 
\\ \midrule
CL-7B    & 38.17          & 28.02          & 25.34        \\
CL-13B   & 38.94          & 21.62          & 20.73        \\
PD-34B   & 49.07          & 37.45          & 32.35        \\
DC-7B    & 49.22          & 29.02          & 23.77         \\
DC-33B   & 45.65          & 32.72          & 29.26          \\ \midrule 
GPT-4    & 52.96          & 40.43 & 31.78 \\ \midrule
Evosuite & 85.71          & 78.91          & 76.59          \\ \bottomrule
\end{tabular}
}
\end{table}







Table~\ref{tab:rq1} shows the effectiveness of each studied open-source LLM with the corresponding best prompt design mentioned above in terms of CSR, $Cov_{L}$, and $Cov_{B}$.
To better understand the effectiveness of open-source LLMs in unit test generation, we also reported the effectiveness of traditional technique Evosuite with its default setting, as well as the closed-source GPT-4 model by incorporating all studied code features in the natural language description style (the most widely-used setting in the literature~\cite{chattester,chatunitest,tang2023chatgpt}). 

From Table~\ref{tab:rq1}, the two most large-scale open-source LLMs (i.e., PD-34B and DC-33B) achieve higher code coverage than the remaining small-scale LLMs, and the commercial GPT-4 generally performs better than these studied open-source LLMs.
For example, DC-33B achieves 12.75\% higher line coverage and 23.10\% higher branch coverage than DC-7B.
This is as expected since larger-scale LLMs tend to have a stronger ability to understand instructions, and the same finding has been observed on the task of LLM-based code generation~\cite{humaneval}. 
There is an unusual case, i.e., CL-13B performs worse than CL-7B in our study. 
This is because CL-7B suffers from a repetition issue~\cite{DBLP:conf/iclr/HoltzmanBDFC20}, which causes that the number of generated unit tests is substantially larger than that generated by CL-13B. 
Although this repetition issue could benefit the task of unit test generation to some degree, it is actually widely-recognized as a typical aspect reflecting LLM's poor capability~\cite{DBLP:journals/corr/abs-1909-05858}.
There are also different findings derived from different tasks with LLMs.
From Table~\ref{tab:rq1}, PD-34B outperforms DC-33B in terms of both line coverage and branch coverage, but the conclusion is opposite on the task of code generation based on the widely-known leaderboard~\cite{bigcode-leaderboard}. 
This indicates the necessity of re-evaluating LLMs when a new task is specified, since the most suitable LLM may be different for different tasks due to different characteristics involved in them.

\finding{While these open-source LLMs have been evaluated on other tasks, not all findings can be generalized to unit test generation, especially in identifying the most effective open-source LLM for it. 
The state-of-the-art commercial GPT-4 is still more effective in unit test generation.}

By comparing the traditional Evosuite with LLM-based unit test generation in Table~\ref{tab:rq1}, the former significantly outperforms all the LLM-based techniques (including the commercial GPT-4) in terms of all three used metrics.
For example, the line coverage achieved by Evosuite is 78.91\%, while that by GPT-4 is just 40.43\%.
Of course, the unit tests generated by Evosuite have always been criticized for poor readability, rather than for low test coverage~\cite{chattester,a3test}.
The poor readability of traditional techniques is also the initial motivation for LLM-based unit test generation~\cite{chattester,tang2023chatgpt,TestPilot}, but definitely, LLM-based unit test generation should not harm test coverage too much compared to Evosuite.
One possible reason is that a substantial portion of the unit tests generated by LLMs (ranging from 34.44\% to 61.78\%) are syntactically invalid.

To potentially improve the effectiveness of LLM-based unit test generation, we further analyzed the underlying root cause of their low CSR.
Specifically, there are three main types of invalid unit tests generated by these LLMs: unresolved symbol errors, parameter mismatch errors, and abstract instantiation errors.
All of them are caused by the \textit{hallucination} issue~\cite{exploringhallucination,Repiolt,DBLP:conf/nips/AgrawalKGLR23} of LLMs.
Unresolved symbol errors are the most common, which account for 30.68\% of all invalid unit tests and are caused due to using semantically-coherent but undefined identifiers in unit tests.
Parameter mismatch errors are the second most common, accounting for 17.25\%, which involves incorrect parameter types or counts in invoked APIs.
This is because there are many APIs sharing the same name but different parameter lists due to the overloading feature, aggregating the difficulty in generating correct API invocations for LLMs.
Abstract instantiation errors are the third most common, accounting for 10.38\%, which involves incorrect instantiation of abstract classes.
Specifically, when a method parameter is defined in an abstract class, an instance of the corresponding subclass should be declared since abstract classes cannot be instantiated. However, LLMs often declare an instance for the parameter according to its defined type.

\finding{LLM's hallucination leads to three main types of compilation errors for unit tests, i.e., unresolved symbol errors, parameter mismatch errors, and abstract instantiation errors.
This makes LLM-based unit test generation underperform Evosuite in terms of test coverage, although the latter suffers from the poor readability issue.}




\begin{table*}[t]
\centering
\caption{Effectiveness of Open-source LLMs with Different In-Context Learning Methods in Terms of Test Coverage.}
\label{tab:rq3_icl_coverage}
\setlength\tabcolsep{3pt}
\begin{tabular}{@{}l|rr|rr|rr|rr|rr@{}}
\toprule
\multicolumn{1}{c|}{\multirow{2}{*}{\textbf{Model}}} & \multicolumn{2}{c|}{\textbf{CL-7B}} & \multicolumn{2}{c|}{\textbf{CL-13B}} & \multicolumn{2}{c|}{\textbf{PD-34B}} & \multicolumn{2}{c|}{\textbf{DC-7B}} & \multicolumn{2}{c}{\textbf{DC-33B}} 
\\ 
 & \textbf{$\bm{Cov_{L}}$}           & \textbf{$\bm{Cov_{B}}$} & \textbf{$\bm{Cov_{L}}$}  & \textbf{$\bm{Cov_{B}}$} & \textbf{$\bm{Cov_{L}}$}  & \textbf{$\bm{Cov_{B}}$} & \textbf{$\bm{Cov_{L}}$}  & \textbf{$\bm{Cov_{B}}$} & \textbf{$\bm{Cov_{L}}$}  & \textbf{$\bm{Cov_{B}}$} 
 \\ \midrule
Base &
  28.02 &
  25.34 &
  21.62 &
  20.73 &
  37.45 &
  32.35 &
  29.02 &
  23.77 &
  32.72 &
  29.26 
  \\


+ CoT 

& {\textbf{-3.04}}
& {-2.79}
& {\textbf{-6.45}}
& {\textbf{-4.57}}
& {-0.02}
& {-2.06}
& {+2.72}
& {\textbf{+4.03}}
& {+0.69}
& {+0.48}

  \\

+ RAG 
& {\textbf{-5.57}}
& {\textbf{-5.85}}
& {\textbf{-6.03}}
& {\textbf{-5.10}}
& {\textbf{-9.28}}
& {\textbf{-8.78}}
& {\textbf{-5.80}}
& {-4.44}
& {-3.34}
& {\textbf{-5.72}}

  \\ 
  \bottomrule

\end{tabular}
\end{table*}
\subsection{Effectiveness of In-Context Learning}
\label{sec:rq3}
We investigated the influence of in-context learning on LLM-based unit test generation by evaluating each studied open-source LLM with the CoT and RAG methods (introduced in Section~\ref{sec:icl}), respectively.
Table~\ref{tab:rq3_icl_coverage} presents their effectiveness in terms of $Cov_L$ and $Cov_B$. We also used bold to highlight the increment/decrement
with statistical significance (p-values smaller than 0.05 and effect
size greater than 0.3).

From this table, incorporating CoT weakens the effectiveness of all three CodeLlama models (i.e., CL-7B, CL-13B, and PD-34B) in terms of $Cov_L$ and $Cov_B$, but enhances the effectiveness of both DeepSeek-Coder models (i.e., DC-7B and DC-33B).
For example, incorporating CoT results in 3.04\% decrement for CL-7B in terms of $Cov_L$, but brings 2.72\% increment for DC-7B.
This difference may be attributed to the ability of LLMs in code comprehension.
Specifically, CoT introduces an extra task of describing the focal method for guiding subsequent unit test generation, which stresses the code comprehension ability of LLMs more explicitly.
Therefore, the LLMs with a strong code comprehension ability could provide informative descriptions, thereby improving the effectiveness of subsequent unit test generation.
However, the LLMs with a relatively weak code comprehension ability may not describe the focal method correctly, eventually leading to the effectiveness decrement.
Through our manual analysis, the studied DeepSeek-Coder models indeed provide more accurate and informative descriptions than the studied CodeLlama models, thereby achieving better effectiveness in terms of test coverage.
This indicates the superiority of the former in the code comprehension ability, which is also aligned with Finding 3.
Hence, it is important to determine whether incorporating CoT to guide the used LLM is really helpful through a preliminary study in advance.

\finding{CoT is helpful to improve the effectiveness of both DeepSeek-Coder models in unit test generation while it \ins{does not benefit}\del{cannot for} the three CodeLlama models due to the stronger ability of the former in code comprehension.}

From Table~\ref{tab:rq3_icl_coverage}, incorporating RAG makes all studied open-source LLMs perform worse in terms of both line coverage and branch coverage.
For example, the decrement of $Cov_L$ and $Cov_B$ for CL-7B is 5.57\% and 5.85\%, respectively. 
We extensively analyzed the reason behind the negative conclusion, and found that there is a significant gap between the unit tests retrieved by the RAG method and those that LLMs excel at generating, which causes the information that LLMs learn from the retrieved examples is limited (even disruptive).
For example, the average length of the retrieved unit tests is 12.10 lines of code (LOC) while that of the generated unit tests by LLMs is 5.60 LOC.
Also, the average number of retrieved unit tests per focal method (i.e., 2.41) is much smaller than that of the generated unit tests by LLMs (i.e., 6.94).
That is, the current retrieval method, which is restricted within the project following the practice of code generation~\cite{rag3, DPR}, is not sufficient.
This indicates that setting a high-quality database specific to the task of unit test generation could be helpful to improve the effectiveness of RAG in enhancing LLM-based unit test generation.



\finding{The RAG method adapted from code generation is ineffective to enhance LLMs in unit test generation.
This is mainly caused by the large gap between unit tests retrieved and those that LLMs excel at generating, indicating the necessity of setting a high-quality retrieval database specific to unit test generation.}

\begin{table}[t]
\centering
\setlength\tabcolsep{3pt}
\caption{Defect Detection Effectiveness of Studied LLMs.}
\label{tab:rq4}
\resizebox{0.95\columnwidth}{!}{
\begin{tabular}{@{}ccccccc@{}}
\toprule
\textbf{Models}      & \textbf{CL-7B} & \textbf{CL-13B} & \textbf{PD-34B} & \textbf{DC-7B} & \textbf{DC-33B} & \textbf{GPT-4} \\ \midrule

NTD  & 41    & 28     & 63     & 60    & 62     & 65    \\
NDD & 12    & 15     & 30      & 24     & 33      & 39    \\ \bottomrule
\end{tabular}
}
\end{table}
\subsection{Defect Detection Ability}
\label{sec:rq4}

We further investigated the effectiveness of each LLM-based unit test generation technique in terms of defect detection by executing its generated unit tests on the corresponding faulty versions.
As shown before, a number of generated unit tests by each LLM cannot be compiled successfully, causing that there are no valid unit tests for many defects.
These defects cannot be detected at all.
Therefore, for each LLM, we filtered out such defects in this experiment and reported the number of remaining defects that are detectable in Row ``NTD'' (short for the \textbf{n}umber of \textbf{t}estable \textbf{d}efects that have at least test classes compiled successfully on the corresponding version) in Table~\ref{tab:rq4}.
From this row, 87.13\% of defects cannot be detected due to the compilation issue on average across all studied LLMs, suggesting that improving the validity of generated unit tests by LLMs is critical to improve the defect detection ability of them.

We then reported the number of detected defects by each LLM in \ins{the row}\del{Row} ``NDD'' in Table~\ref{tab:rq4}.
We further confirmed the larger-scale LLMs (including GPT-4, PD-34B, and DC-33B) detected more defects than the remaining smaller-scale LLMs. \ins{Among the detected defects, there are two types of defect-triggering reasons: run-time exceptions and assertion errors. For example, DC-7B, PD-34B, DC-33B, and GPT-4 detected 6, 16, 17, and 19 defects triggered by assertion errors, respectively, and 18, 14, 16, and 20 defects triggered by run-time exceptions. On the other hand, all defects detected by CL-7B and CL-13B were triggered by run-time exceptions, showing their relatively weaker ability in generating proper assertions.}
\ins{Despite these findings,}\del{Among these detectable defects, however,} the percentage of detected defects is still small, ranging from 29.27\% to 60.00\%.
This indicates the weak defect detection ability of LLM-based unit test generation.



\finding{The defect detection ability of LLM-based unit test generation is weak, mainly caused by the low validity of generated unit tests.
On average, for 87.13\% defects, LLMs do not generate any valid unit tests. 
Among the remaining defects, only 47.28\% defects are exactly detected. }


We then analyzed why just a few defects are detected among all detectable ones in order to further understand the challenges for LLM-based unit test generation in defect detection.
\ins{Specifically, following the PIE theory in software engineering~\cite{PIE_Theory}, we designed three reasons why these defects were not detected by generated tests. Then, four authors with over four years of Java programming experience manually analyzed and independently labeled each defect. The Cohen's Kappa score is 0.95, indicating the labeling quality to some degree. Then, all the inconsistencies in labeling were discussed together to reach a consensus.}\del{Through our manual investigation of each valid unit test corresponding to each undetected defect, we summarized the following three reasons.}
Table~\ref{tab:rq4-reason} presents the number of undetected defects caused by each reason for each LLM.
\begin{table*}[t]
\centering
\caption{Number of Defects that Valid Unit Tests Failed to Detect Categorized by Three Reasons.}
\label{tab:rq4-reason}

\begin{tabular}{lcccccc}
\toprule
\textbf{Model} &
  \textbf{CL-7B} &   
  \textbf{CL-13B} &
  \textbf{PD-34B} &
  \textbf{DC-7B} &
  \textbf{DC-33B} &
  \textbf{GPT-4} 
  \\ \midrule
Insufficient Test Coverage   & 6  & 3  & 11 & 10 & 9  & 0  
\\
Missing Specific Inputs   & 22 & 10  & 21 & 26 & 20 & 24 
\\
Improper Assertions    & 1  & 0  & 1  & 0  & 0  & 2  
\\ 
\bottomrule
\end{tabular}%

\end{table*}



(1) \textbf{Insufficient test coverage} refers to that all the generated unit tests by an LLM do not reach the faulty code. 
From Table~\ref{tab:rq4-reason}, there are 3.61\%, 1.81\%, 6.63\%, 6.02\%, 5.42\%
of undetected defects caused by this reason for CL-7B, CL-13B, PD-34B, DC-7B, \ins{and }DC-33B, respectively.
(2) \textbf{Missing specific inputs} refers to that there are some unit tests executing the faulty code but the inputs specified in the unit tests cannot trigger the defect.
That is, the triggering of the defect requires some specific inputs (such as setting the real part of a complex number to {\tt NaN} for triggering the defect in Math-53).
From Table~\ref{tab:rq4-reason}, there are 13.25\%, 6.02\%, 12.65\%, 15.66\%, 12.05\% , 14.46\%
of undetected defects caused by this reason for CL-7B, CL-13B, PD-34B, DC-7B, DC-33B, \ins{and }GPT-4, respectively.
(3) \textbf{Improper assertions} refer to that the defect-triggering inputs have been specified in some unit tests, but no proper assertions are produced to capture this defect.
For example, the defect in Compress-34 lies in missing to flush the input stream after closing a file. 
However, the unit tests generated by PD-34B do not contain the assertions checking whether the corresponding attribute is {\tt null}, but just check whether the file is closed.
From Table~\ref{tab:rq4-reason}, there are 0.60\%, 0.60\%, 1.20\%
of undetected defects caused by this reason for CL-7B, PD-34B, \ins{and }GPT-4, respectively.
\finding{Regarding the undetected defects with valid unit tests generated by LLMs, the main reason lies in missing to produce the specific inputs required to trigger defects, resulting in 74.99\% of undetected defects on average.}

\section{Threats to Validity}
\label{sec:threats}
The threats to \textit{internal} validity mainly lie in our experimental scripts and the usages of LLMs.
To reduce the former threat, two authors have carefully checked our implementations through code review and testing.
We also released our implementations at our project homepage for replication.
To reduce the latter threat, we used publicly accessible model weights from Hugging Face adhered strictly to the well-documented usage guidelines for open-source models, and leveraged the API services provided by OpenAI for GPT-4.

The threats to \textit{external} validity mainly lie in our used benchmark and LLMs.
In our study, we adopted the Defects4J benchmark for evaluation, which has been widely used in the existing studies on software testing and debugging~\cite{xia2023automated,jiang2023variable,liang2021interactive,jiang2019combining}.
Specifically, we collected 778 focal methods involving 413 defects from 17 projects, exhibiting the diversity of our evaluation data to some degree.
Also, we studied five open-source LLMs and the state-of-the-art commercial LLM (i.e., GPT-4). 
As presented in Section~\ref{sec:LLMs}, we selected these LLMs by considering diverse architectures and model sizes, as well as their effectiveness on code-related tasks highlighted in the leaderboard hosted on Hugging Face.
The above diversity is helpful to reduce this kind of threats to some degree.
Actually, the current experiments on the used benchmark and LLMs have spent 3,000 A100 GPU hours, and our experimental design balances conclusion generalizability and evaluation costs well. 




The threats to \textit{construct} validity mainly lie in our studied code features, ICL methods, randomness, and potential data leakage.
Regarding the former, we collected all the code features that have been used in the existing studies on LLM-based testing~\cite{chattester,chatunitest,a3test,athenatest}, demonstrating the comprehensiveness of our investigation.
However, due to the evaluation costs, it is unaffordable to investigate the effectiveness of all combinations of these code features.
Instead, we performed an ablation experiment by respectively removing one code feature from the whole set to investigate the contribution.
Therefore, we may not find the \textit{globally} optimal setting for each studied LLM for the experiments in Sections~\ref{sec:rq2},~\ref{sec:rq3}, and~\ref{sec:rq4}, but selected the \textit{locally} optimal setting from our ablation experiment.

Regarding the ICL methods, we investigated two most widely-studied methods, i.e., CoT and RAG, in the literature~\cite{rag,CoT_scaling}.
According to the characteristics of unit test generation, we adapted them to fit our task as introduced in Section~\ref{sec:icl}.
The current adaptations are reasonable but may be not the best strategies in unit test generation.
In the future, we will explore more usages of CoT, RAG, and some other ICL methods (e.g., self-repairing~\cite{chatunitest}, self-consistency~\cite{self-consistency}, prompt tuning~\cite{prompt-tuning}) in LLM-based unit test generation for more sufficient investigation.
To reduce the threat of randomness from LLMs, we set the temperature to zero, which is the most widely-used setting to increase LLMs' determinacy~\cite{DBLP:journals/corr/abs-2308-02828,DBLP:conf/fose-ws/FanGHLSYZ23}.
Same as the existing studies on LLMs using the Defects4J benchmark~\cite{chatunitest,DBLP:journals/corr/abs-2312-14898,DBLP:journals/corr/abs-2210-14179}, there is also a potential data leakage threat in our study.
Following the existing practice~\cite{TestPilot,DBLP:journals/corr/abs-2210-14179},
we compared LLM-generated unit tests with the original unit tests equipped by this benchmark.
We found that there is no exact match between them, and even the number of LLM-generated unit tests (3.70 on average) is largely different with that of original unit tests (2.41).
This helps reduce the influence of this threat to some degree.
In the future, we will extend our study to more recent benchmarks, such as GitBug-Java~\cite{gitbug}.
\section{Implications}

Based on our findings from the study, we summarize a series of implications, which could suggest better practice of leveraging LLMs in unit test generation.


\textbf{\textcircled{1} Tuning prompt design (including both description styles and code features) for a given LLM is important.} As shown in Section~\ref{sec:rq1}, both description styles and code features significantly affect the effectiveness of LLM-based unit test generation. Improper prompts can even decrease effectiveness in terms of both CSR and test coverage. This suggests that, for a given LLM, it is necessary to perform preliminary experiments to determine an optimal prompt for unit test generation.

\textbf{\textcircled{2} Removing useless information from code features helps balance prompt content and space for generating unit tests.} According to Section~\ref{sec:rq1}, there is a trade-off between incorporating more information and leaving sufficient space for generating unit tests, particularly for $FC_m$. Directly removing $FC_m$ is somewhat coarse-grained and may result in the loss of necessary information for constructing valid unit tests. Therefore, conducting static analysis to remove the methods in $FC_m$ that do not have dependencies to the focal method is a promising method to achieve this balance.




\textbf{\textcircled{3} Empirically selecting a proper LLM is necessary, instead of relying on the best LLM according to experience in other tasks.} As shown in Section~\ref{sec:rq2}, larger-scale LLMs generally perform better than smaller-scale ones, a conclusion consistent across unit test generation and other code-related tasks. However, for LLMs with similar scales, the best one may vary depending on the specific task. This variability arises because open-source LLMs are based on different foundation models and fine-tuned on different datasets, thus exhibiting different capabilities across various tasks. Therefore, to generate high-quality tests, it is crucial to evaluate several LLMs rather than simply using the one deemed best in other tasks.

\textbf{\textcircled{4} Refining the use of ICL methods specific to unit test generation is essential.} As shown in Section~\ref{sec:rq3}, the effectiveness of CoT varies due to different LLMs' code comprehension abilities, which significantly affects their understanding of focal methods and, consequently, the effectiveness of unit test generation. Specifically, applying CoT to LLMs with weak code comprehension abilities can even decrease test coverage. Therefore, it is advisable to conduct preliminary studies to determine if employing CoT benefits the given LLM.
Additionally, the existing RAG method used in code generation is ineffective for unit test generation, primarily due to the significant gap between retrieved unit tests and those generated by LLMs. This highlights the necessity of constructing a high-quality retrieval database specific to unit test generation.

\textbf{\textcircled{5} Designing post-processing strategies to fix invalid unit tests can help mitigate the influence of LLMs' hallucination.} Our study reveals that a significant percentage of unit tests generated by LLMs cannot be compiled successfully, thereby limiting their effectiveness in terms of both test coverage and defect detection. We analyzed the major error types for these invalid unit tests caused by LLMs' hallucination, and found that it is possible to design post-processing rules to fix these errors. For example, extracting the class inheritance relationship can help address abstract instantiation errors. Additionally, we can follow existing methods~\cite{chattester,chatunitest} to instruct LLMs to fix these errors by feeding error messages back to them. 

\textbf{\textcircled{6} Designing mutation strategies for the inputs specified in LLM-generated unit tests is helpful in improving defect detection ability.} As demonstrated in Section~\ref{sec:rq4}, many detectable defects remain undiscovered by valid unit tests generated by LLMs due to the absence of specific defect-triggering inputs. This is understandable as LLMs tend to produce more common inputs, given their inherent characteristics. Therefore, it could be promising to design mutation strategies to transform a generated unit test into a mutated one that includes the defect-triggering input. Such mutation strategies could involve (1) mutating an input to its boundary value, and (2) mutating an input to a value found in historically defect-triggering tests, which can be obtained from bug repositories.

\textbf{\textcircled{7} Supervised fine-tuning (SFT) of open-source LLMs by incorporating unit test generation data may fundamentally improve effectiveness.} The overall effectiveness of directly applying existing open-source LLMs to the task of unit test generation is unsatisfactory based on our results. This deficiency may stem from these LLMs' limited knowledge acquisition regarding this task during the pre-training process. Therefore, conducting SFT specific to the task of unit test generation could be beneficial, representing a promising direction alongside the construction of a high-quality corpus tailored to this task.
\section{Related Work}
\label{sec:related_work}

\noindent\textbf{Unit Test Generation.} In the literature, a significant body of research has been dedicated to the development of automated unit test generation techniques~\cite{serra2019effectiveness,athenatest,a3test,chattester,mutap,chatunitest,exploreingllminutg,DBLP:journals/corr/abs-2312-10622,DBLP:journals/corr/abs-2404-13340,TELPA}. They can be divided into two main categories: traditional techniques and DL-based techniques.

Traditional techniques adopted a variety of methods to solve this problem, including symbolic execution \cite{DBLP:conf/tacas/XieMSN05,DBLP:conf/issta/PasareanuMBGLPP08}, search-based optimization \cite{evosuite}, model checking \cite{DBLP:journals/sttt/EnoiuCOWSP16,DBLP:conf/esec/GargantiniH99}.
For instance, one of the most influential unit test generation techniques, i.e., Evosuite \cite{evosuite}, treats this problem as a search problem and leverages the evolutionary search algorithm to find the tests achieving high code coverage.
Although they can generate unit tests with reasonable coverage, they often lack readability and maintainability compared to manually written ones. As a result, it is challenging for developers to use these automatically generated tests in practice directly.

DL-based techniques exploited the power of pre-trained language models to unit test generation. 
They often treat this problem as a neural machine translation problem, which inputs the target methods and outputs the unit tests. 
For instance, Tufano et al.~\cite{athenatest} proposed AthenaTest, which leverages the pre-trained BART Transformer~\cite{bart} to generate unit tests about the given methods.
However, the early techniques used smaller general-purpose pre-trained language models, thereby limiting their effectiveness.

Recently, with the huge impact of ChatGPT, researchers explored the effectiveness of LLMs in generating unit tests.
Specifically, \ins{Yuan}\del{Zhang} et al.~\cite{chattester} proposed ChatTester that iteratively generates unit tests based on conversations with ChatGPT.
Lemieux et al.~\cite{lemieux2023codamosa} proposed CODAMOSA that combines evolutionary search with CodeX~\cite{humaneval} by exploiting code understanding ability of CodeX to overcome the low rate of generating coverage-improving tests with traditional methods.
Li et al.~\cite{li2023nuances} proposed using the subtle differences between fixed and buggy code versions to guide ChatGPT to generate failure-inducing tests.
\ins{Schafer et al.~\cite{TestPilot} proposed TestPilot, which incorporates API documentation and GPT-3.5 for unit test generation.}
These LLM-based techniques show significant superiority over earlier DL-based ones.
However, they are all based on closed-source LLMs, resulting in a lack of transparency and reproducibility.
\ins{Recently, Yang et al.~\cite{TELPA} proposed TELPA, which aims to improve the effectiveness of LLMs in unit test generation by integrating them with bidirectional method invocation analysis and counterexample sampling. However, their evaluation was limited to PD-34B, and it remains unclear how recent open-source LLMs perform in unit test generation.}

\del{Particularly, it is still unclear how recent open-source LLMs perform in unit test generation.}

\noindent\textbf{Empirical Study on evaluating LLMs in Code-Related Tasks.}
With the rapid development in LLMs, there are multiple studies investigating the effectiveness of LLMs in code-related tasks~\cite{ClassEval,nam2024using,DBLP:journals/corr/abs-2305-12865,molina2024test,DBLP:journals/corr/abs-2210-14179,DBLP:conf/icse/NiuLNCGL23,DBLP:journals/tse/WangHCLWW24,pan2024lost,li2024extracting,evalplus,DBLP:journals/corr/abs-2308-10620,mastropaolo2023robustness,li2023exploring,guo2024exploring,chen2024deep}. 
In particular, there are some empirical studies on LLM-based unit test generation~\cite{chattester,chatunitest,exploreingllminutg,TestPilot,mutap}. 
For example, Yuan et al.~\cite{chattester} investigated the effectiveness of ChatGPT \ins{in terms of the correctness, sufficiency, readability and usability of generated unit tests in unit test generation.}\del{and proposed ChatTester by employing CoT for unit test generation.}
\del{Xie et al.~\cite{chatunitest} proposed and evaluated ChatUniTest by employing the feedback-based generation process with ChatGPT.}
Siddiq et al.~\cite{exploreingllminutg} investigated the effectiveness of GPT-3.5 and Codex in unit test generation. 
Schafer et al.~\cite{TestPilot} \ins{investigated the effectiveness of GPT-3.5 in generating unit tests for Javascript language.}\del{proposed TestPilot, which incorporates API documentation and GPT-3.5 for unit test generation}.

However, these existing studies mostly relied on fixed prompting strategies based on closed-source LLMs, neglecting the power of advanced open-source LLMs and the influence of various prompting factors (such as prompt design and ICL methods). 
In our study, we focus on open-source LLMs with extensive consideration of these factors, and performed the first study to demonstrate their effectiveness. 
Our findings provide a series of actionable suggestions to LLM-based unit test generation.

Furthermore, there are some empirical studies on evaluating LLMs in addressing other code-related tasks.
For example, Du et al.~\cite{ClassEval} conducted experiments to investigate the effectiveness of LLMs in class-level code generation. Nam et al.~\cite{nam2024using} compared the code understanding ability of LLMs when using them to help developers understand the project. Sun et al.~\cite{DBLP:journals/corr/abs-2305-12865} conducted experiments to investigate the effectiveness of ChatGPT in code summarization. 
Facundo et al.~\cite{molina2024test} summarized three types of oracles and threats that arise from automatically generating oracles using LLMs.
Researchers~\cite{DBLP:journals/corr/abs-2210-14179,jiang2024evaluating} conducted experiments on LLMs to investigate their effectiveness in automated program repair. 
Niu et al.~\cite{DBLP:conf/icse/NiuLNCGL23} conducted a comprehensive investigation on the effectiveness of pre-trained models in six code-related tasks.
Different from them, our work targets the task of unit test generation.

\ins{
\noindent\textbf{Test Oracle Generation}
Due to the importance of test oracles, numerous research efforts have been dedicated to automatically generating test oracles for test cases~\cite{dinella2022toga,ATLAS,Yu2022IRbased,EDITAS}. For example, Watson et al.~\cite{ATLAS} proposed ATLAS, which adopted a bidirectional RNN for test assertion generation. Dinella et al.~\cite{dinella2022toga} introduced TOGA to infer exceptional and assertion test oracles based on the context of the focal method. Yu et al.~\cite{Yu2022IRbased} first proposed an IR-based method for test assertion generation, and Sun et al.~\cite{EDITAS} further developed EDITAS to improve IR-based test assertion generation by considering the differences between the retrieved and target ``focal-test'' method pairs.
In our study, we use LLMs to generate complete unit tests, including both test prefixes and test oracles. When measuring the effectiveness of generated unit tests (e.g., validity and defect-detection effectiveness), we also considered both test prefixes and oracles.}

\section{Conclusion}
This work conducted an comprehensive evaluation of five open-source LLMs for unit test generation. We investigated several aspects: the influence of different prompt designs, the comparison effectiveness among open-source LLMs, the commercial closed-source LLM (i.e., GPT-4) and the traditional method (i.e., Evosuite), the influence of in-context learning on LLM-based unit tests generation, and the defect detection ability of LLM-generated unit tests.
Our findings first point out that prompt design is crucial to the effectiveness of LLMs in unit test generation, and suggest that it is better to align the description style with the training data and choose code features considering the LLMs' code comprehension ability and the prompt length.
Second, our results indicate that the conclusions drawn from other tasks do not necessarily generalize to unit test generation, especially in identifying the best open-source LLM specific to the task. 
However, all studied LLMs (including the best GPT-4) still underperform Evosuite, primarily due to the significant percentage of invalid tests caused by hallucination. Our further analysis suggests that effective post-processing could help mitigate this problem.
Third, we found that directly adapting ICL techniques from other tasks does not improve the effectiveness of unit tests generation, requiring special design for the use of ICL methods according to the characteristics of the task.
Finally, we found that the defect detection ability of LLM-generated unit tests is limited, primarily due to their low syntactic validity and the lack of specific input for triggering the defects. Our analysis indicates that designing effective mutation strategies for the inputs could further improve the defect detection effectiveness.

\begin{acks}
\ins{We gratefully acknowledge the financial support provided by the   National Natural Science Foundation of China (Grant Nos. 62322208, 62202040, 62232001, 12411530122) and CCF-Huawei Populus Grove Fund which made this research possible.
We would like to express our sincere appreciation to the anonymous reviewers for their constructive suggestions, which have greatly improved the quality of this paper.}
\end{acks}

\balance
\bibliographystyle{ACM-Reference-Format}
\bibliography{ref}

\end{document}